\documentclass[runningheads,a4paper]{llncs}
\usepackage{times}
\usepackage{color}
\usepackage{latexsym}
\usepackage{amsmath}
\usepackage{amssymb}
\usepackage{graphicx}
\usepackage{multirow}
\usepackage{makecell}
\usepackage{stfloats}

\usepackage{url}
\usepackage{enumitem}
\setenumerate[1]{itemsep=2pt,partopsep=0pt,parsep=\parskip,topsep=3pt,leftmargin=13pt}
\newcommand{\tabincell}[2]{\begin{tabular}{@{}#1@{}}#2\end{tabular}}
\begin{document}

\mainmatter  % start of an individual contribution

\title{QA4IE: A Question Answering based Framework\\
for Information Extraction}

\author{Lin Qiu\inst{1}\and Hao Zhou\inst{1}\and Yanru Qu\inst{1}\and Weinan Zhang\inst{1}\and
Suoheng Li\inst{2}\and\\ Shu Rong\inst{2}\and Dongyu Ru\inst{1}\and
Lihua Qian\inst{1}\and Kewei Tu\inst{3}\and Yong Yu\inst{1}}
\authorrunning{L. Qiu et al.}

\institute{
Shanghai Jiao Tong University \\
\email{\{lqiu, kevinqu, maxru, qianlihua, yyu\}@apex.sjtu.edu.cn \\
\{zhou1998, wnzhang\}@sjtu.edu.cn}\and
Yitu Tech\\
\email{\{suoheng.li, shu.rong\}@yitu-inc.com}\and
ShanghaiTech University\\
\email{tukw@shanghaitech.edu.cn}
}


\newcommand{\qu}[1]{{\bf \color{blue} [qyr ``#1'']}}

\maketitle

\begin{abstract}
Information Extraction (IE) refers to automatically extracting structured relation tuples from unstructured texts. Common IE solutions, including Relation Extraction (RE) and open IE systems, can hardly handle cross-sentence tuples, and are severely restricted by limited relation types as well as informal relation specifications (e.g., free-text based relation tuples). In order to overcome these weaknesses, we propose a novel IE framework named QA4IE, which leverages the flexible question answering (QA) approaches to produce high quality relation triples across sentences. Based on the framework, we develop a large IE benchmark with high quality human evaluation. This benchmark contains 293K documents, 2M golden relation triples, and 636 relation types. We compare our system with some IE baselines on our benchmark and the results show that our system achieves great improvements.
\end{abstract}

\section{Introduction and Background}
Information Extraction (IE), which refers to extracting structured information (i.e., relation tuples) from unstructured text, is the key problem in making use of large-scale texts.
High quality extracted relation tuples can be used in various downstream applications such as Knowledge Base Population \cite{ji2011knowledge}, Knowledge Graph Acquisition \cite{lin2015learning}, and Natural Language Understanding. However, existing IE systems still cannot produce high-quality relation tuples to effectively support downstream applications.

\subsection{Previous IE Systems}
Most of previous IE systems can be divided into Relation Extraction (RE) based systems \cite{mintz2009distant,zeng2015distant} and Open IE systems \cite{angeli2015leveraging,del2013clausie,schmitz2012open}.

Early work on RE decomposes the problem into Named Entity Recognition (NER) and relation classification. With the recent development of neural networks (NN), NN based NER models \cite{lample2016neural,ma2016end} and relation classification models \cite{xu2015semantic} show better performance than previous handcrafted feature based methods. The recently proposed RE systems \cite{ren2017cotype,zheng2017joint} try to jointly perform entity recognition and relation extraction to improve the performance.
One limitation of existing RE benchmarks, e.g., NYT \cite{riedel2010modeling}, Wiki-KBP \cite{ling2012fine} and BioInfer \cite{pyysalo2007bioinfer}, is that they only involve 24, 19 and 94 relation types respectively comparing with thousands of relation types in knowledge bases such as DBpedia \cite{auer2007dbpedia,bizer2009dbpedia}.
Besides, existing RE systems can only extract relation tuples from a single sentence while the cross-sentence information is ignored. Therefore, existing RE based systems are not powerful enough to support downstream applications in terms of performance or scalability.

On the other hand, early work on Open IE is mainly based on bootstrapping and pattern learning methods \cite{agichtein2000snowball}. Recent work incorporates lexical features and sentence parsing results to automatically build a large number of pattern templates, based on which the systems can extract relation tuples from an input sentence \cite{angeli2015leveraging,del2013clausie,schmitz2012open}.
An obvious weakness is that the extracted relations are formed by free texts which means they may be polysemous or synonymous and thus cannot be directly used without disambiguation and aggregation. The extracted free-text relations also bring extra manual evaluation cost, and how to automatically evaluate different Open IE systems fairly is an open problem.
Stanovsky and Dagan \cite{stanovsky2016creating} try to solve this problem by creating an Open IE benchmark with the help of QA-SRL annotations \cite{he2015question}. Nevertheless, the benchmark only involves 10K golden relation tuples.
Hence, Open IE in its current form cannot provide a satisfactory solution to high-quality IE that supports downstream applications.

There are some recently proposed IE approaches which try to incorporate Question Answering (QA) techniques into IE. Levy et al. \cite{levy2017zero} propose to reduce the RE problem to answering simple reading comprehension questions. They build a question template for each relation type, and by asking questions with a relevant sentence and the first entity given, they can obtain relation triples from the sentence corresponding to the relation type and the first entity. Roth et al. \cite{roth2018neural} further improve the model performance on a similar problem setting. However, these approaches focus on sentence level relation argument extractions and do not provide a full-stack solution to general IE. In particular, they do not provide a solution to extract the first entity and its corresponding relation types before applying QA. Besides, sentence level relation extraction ignores the information across sentences such as coreference and inference between sentences, which greatly reduces the information extracted from the documents.

\subsection{QA4IE Framework}
To overcome the above weaknesses of existing IE systems, we propose a novel IE framework named QA4IE to perform document level general IE with the help of state-of-the-art approaches in Question Answering (QA) and Machine Reading Comprehension (MRC) area.

\begin{figure}[t]
\vskip -0.1in
\begin{center}
\centerline{\includegraphics[width=0.8\columnwidth]{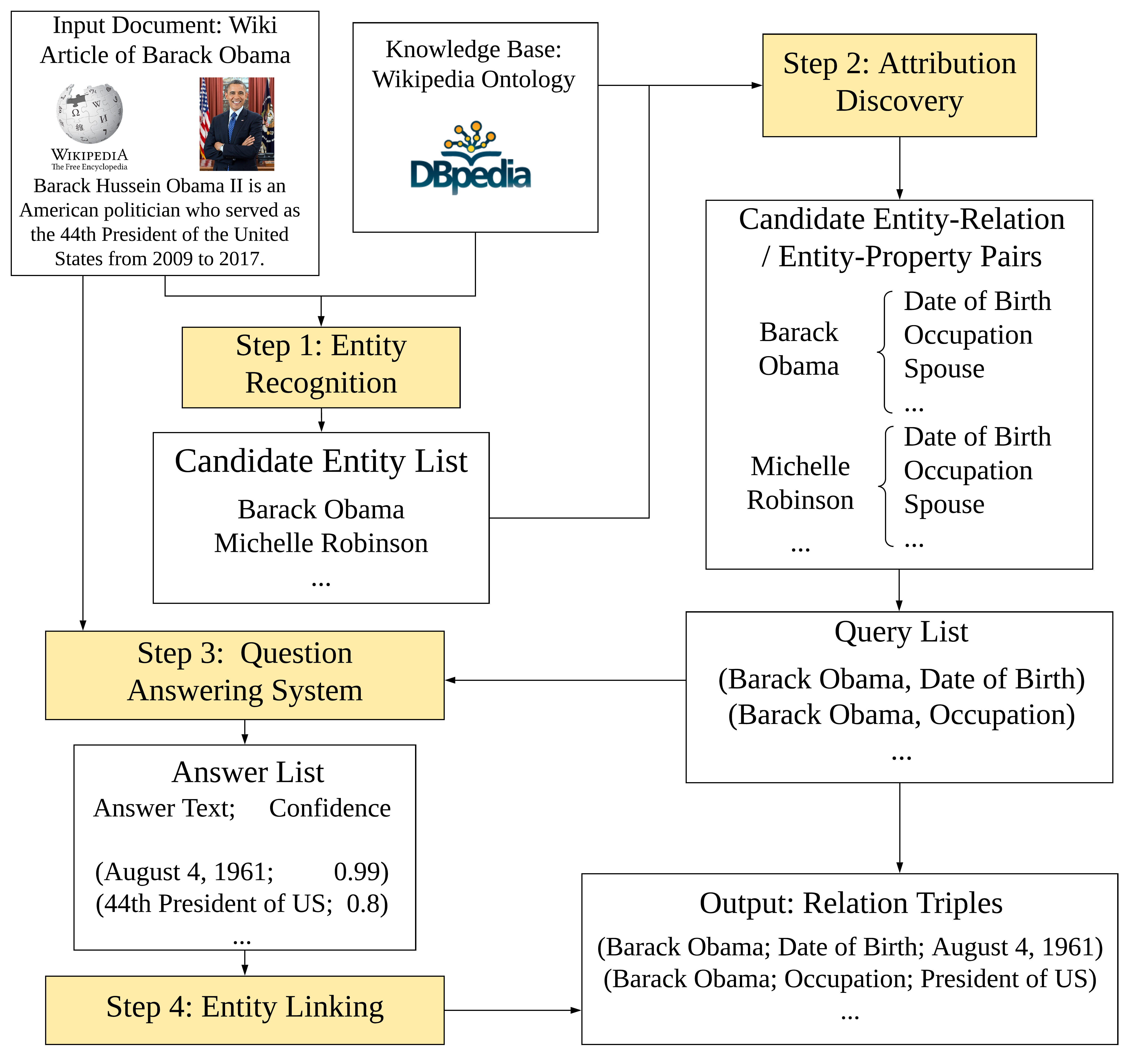}}
\vskip -0.15in
\caption{An overview of our QA4IE Framework.}
\label{framework}
\end{center}
\vskip -0.48in
\end{figure}

The input of QA4IE is a document $D$ with an existing knowledge base $K$ and the output is a set of relation triples $R = \{e_i, r_{ij}, e_j\}$ in $D$ where $e_i$ and $e_j$ are two individual entities and $r_{ij}$ is their relation. We ignore the adverbials and only consider the entity pairs and their relations as in standard RE settings. Note that we process the entire document as a whole instead of processing individual sentences separately as in previous systems.
As shown in Figure \ref{framework}, our QA4IE framework consists of four key steps:
\begin{enumerate}
  \item Recognize all the candidate entities in the input document $D$ according to the knowledge base $K$. These entities serve as the first entity $e_i$ in the relation triples $R$.
  \item For each candidate entity $e_i$, discover the potential relations/properties as $r_{ij}$ from the knowledge base $K$.
  \item Given a candidate entity-relation or entity-property pair $\{e_i, r_{ij}\}$ as a query, find the corresponding entity or value $e_j$ in the input document $D$ using a QA system. The query here can be directly formed by the word sequence of $\{e_i, r_{ij}\}$, or built from templates as in \cite{levy2017zero}.
  \item Since the results of step 3 are formed by free texts in the input document $D$, we need to link the results to the knowledge base $K$.
\end{enumerate}

This framework determines each of the three elements in relation triples step by step. Step 1 is equivalent to named entity recognition (NER), and state-of-the-art NER systems \cite{luo2015joint,ma2016end} can achieve over 0.91 F1-score on CoNLL'03 \cite{tjong2003introduction}, a well-known NER benchmark. For attribution discovery in step 2, we can take advantage of existing knowledge base ontologies such as Wikipedia Ontology to obtain a candidate relation/property list according to NER results in step 1. Besides, there is also some existing work on attribution discovery \cite{lee2013attribute,yahya2014renoun} and ontology construction \cite{gupta2014biperpedia} that can be used to solve the problem in step 2. The most difficult part in our framework is step 3 in which we need to find the entity (or value) $e_j$ in document $D$ according to the previous entity-relation (or entity-property) pair $\{e_i, r_{ij}\}$. Inspired by recent success in QA and MRC \cite{seo2016bidirectional,wang2016machine,wang2017gated}, we propose to solve step 3 in the setting of SQuAD \cite{rajpurkar2016squad} which is a very popular QA task. The problem setting of SQuAD is that given a document $\tilde{D}$ and a question $q$, output a segment of text $a$ in $\tilde{D}$ as the answer to the question. In our framework, we assign the input document $D$ as $\tilde{D}$ and the entity-relation (or entity-property) pair $\{e_i, r_{ij}\}$ as $q$, and then we can get the answer $a$ with a QA model. Finally in step 4, since the QA model can only produce answers formed by input free texts, we need to link the answer $a$ to an entity $e_j$ in the knowledge base $K$, and the entity $e_j$ will form the target relation triple as $\{e_i, r_{ij}, e_j\}$. Existing Entity Linking (EL) systems \cite{moro2014entity,shen2015entity} directly solve this problem especially when we have high quality QA results from step 3.

As mentioned above, step 1, 2 and 4 in the QA4IE framework can be solved by existing work. Therefore, in this paper, we mainly focus on step 3. According to the recent progress in QA and MRC, deep neural networks are very good at solving this kind of problem with a large-scale dataset to train the network. However, all previous IE benchmarks \cite{stanovsky2016creating} are too small to train neural network models typically used in QA, and thus we need to build a large benchmark. Inspired by WikiReading \cite{hewlett2016wikireading}, a recent large-scale QA benchmark over Wikipedia, we find that the articles in Wikipedia together with the high quality triples in knowledge bases such as Wikidata \cite{vrandevcic2014wikidata} and DBpedia can form the supervision we need. Therefore, we build a large scale benchmark named QA4IE benchmark which consists of 293K Wikipedia articles and 2M golden relation triples with 636 different relation types.

Recent success on QA and MRC is mainly attributed to advanced deep learning architectures such as attention-based and memory-augmented neural networks \cite{bahdanau2015neural,sukhbaatar2015end} and the availability of large-scale datasets \cite{hermann2015teaching,hill2015goldilocks} especially SQuAD.
The differences between step 3 and SQuAD can be summarized as follows. First, the answer to the question in SQuAD is restricted to a continuous segment of the input text, but in QA4IE, we remove this constraint which may reduce the number of target relation triples. Second, in existing QA and MRC benchmarks, the input documents are not very long and the questions may be complex and difficult to understand by the model, while in QA4IE, the input documents may be longer but the questions formed by entity-relation (or entity-property) pair are much simpler. Therefore, in our model, we incorporate Pointer Networks \cite{vinyals2015pointer} to adapt to the answers formed by any words within the document in any order as well as Self-Matching Networks \cite{wang2017gated} to enhance the ability on modeling longer input documents.

\subsection{Contributions}
The contributions of this paper are as follows:
\begin{enumerate}
  \item We propose a novel IE framework named QA4IE to overcome the weaknesses of existing IE systems. As we discussed above, the problem of step 1, 2 and 4 can be solved by existing work and we propose to solve the problem of step 3 with QA models.
  \item To train a high quality neural network QA model, we build a large IE benchmark in QA style named QA4IE benchmark which consists of 293K Wikipedia articles and 2 million golden relation triples with 636 different relation types.
  \item To adapt QA models to the IE problem, we propose an approach that enhances existing QA models with Pointer Networks and Self-Matching Networks.
  \item We compare our model with IE baselines on our QA4IE benchmark and achieve a great improvement over previous baselines.
  \item We open source our code and benchmark for repeatable experiments and further study of IE.\footnote{Our source code and benchmark datasets can be found at https://github.com/SJTU-lqiu/QA4IE}
\end{enumerate}

\section{QA4IE Benchmark Construction}
This section briefly presents the construction pipeline of QA4IE benchmark to solve the problem of step 3 as in our framework (Figure \ref{framework}). Existing largest IE benchmark \cite{stanovsky2016creating} is created with the help of QA-SRL annotations \cite{he2015question} which consists of 3.2K sentences and 10K golden extractions. Following this idea, we study recent large-scale QA and MRC datasets and find that WikiReading \cite{hewlett2016wikireading} creates a large-scale QA dataset based on Wikipedia articles and WikiData relation triples \cite{vrandevcic2014wikidata}. However, we observe about 11\% of QA pairs with errors such as wrong answer locations or mismatch between answer string and answer words. Besides, there are over 50\% of QA pairs with the answer involving words out of the input text or containing multiple answers. We consider these cases out of the problem scope of this paper and only focus on the information within the input text.

Therefore, we choose to build the benchmark referring the implementation of WikiReading based on Wikipedia articles and golden triples from Wikidata and DBpedia \cite{auer2007dbpedia,bizer2009dbpedia}. Specifically, we build our QA4IE benchmark in the following steps.

\noindent\textbf{Dump and Preprocessing.} We dump the English Wikipedia articles with Wikidata knowledge base and match each article with its corresponding relation triples according to its title. After cleaning data by removing low frequency tokens and special characters, we obtain over 4M articles and 18M triples with over 800 relation types.

\noindent\textbf{Clipping.}~We discard the triples with multiple entities (or values) for $e_j$ (account for about 6\%, e.g., a book may have multiple authors). Besides, we discard the triples with any word in $e_j$ out of the corresponding article (account for about 50\%). After this step, we obtain about 3.5M articles and 9M triples with 636 relation types.

\noindent\textbf{Incorporating DBpedia.} Unlike WikiData, DBpedia is constructed automatically without human verification. Relations and properties in DBpedia are coarse and noisy. Thus we fix the existing 636 relation types in WikiData and build a projection from DBpedia relations to these 636 relation types. We manually find 148 relations which can be projected to a WikiData relation out of 2064 DBpedia relations. Then we gather all the DBpedia triples with the first entity is corresponding to one of the above 3.5M articles and the relation is one of the projected 148 relations. After the same clipping process as above and removing the repetitive triples, we obtain 
394K additional triples in
302K existing Wikipedia articles.

\noindent\textbf{Distillation.}~Since our benchmark is for IE, we prefer the articles with more golden triples involved by assuming that Wikipedia articles with more annotated triples are more informative and better annotated. Therefore, we figure out the distribution of the number of golden triples in articles and decide to discard the articles with less than 6 golden triples (account for about 80\%). After this step, we obtain about 200K articles and 1.4M triples with 636 relation types.

\noindent\textbf{Query and Answer Assignment.}~For each golden triple $\{e_i, r_{ij}, e_j\}$, we assign the relation/property $r_{ij}$ as the query and the entity $e_j$ as the answer because the Wikipedia article and its corresponding golden triples are all about the same entity $e_i$ which is unnecessary in the queries. Besides, we find the location of each $e_j$ in the corresponding article as the answer location. As we discussed in Section 1, we do not restrict $e_j$ to a continuous segment in the article as required in SQuAD. Thus we first try to detect a matched span for each $e_j$ and assign this span as the answer location. Then for each of the rest $e_j$ which has no matched span, we search a matched sub-sequence in the article and assign the index sequence as the answer location. We name them \textbf{span-triples} and \textbf{seq-triples} respectively. Note that each triple will have an answer location because we have discarded the triples with unseen words in $e_j$ and if we can find multiple answer locations, all of them will be assigned as ground truths.

\begin{table}[!t]
\centering
\vskip -0.05in
\caption{Detailed Statistics of QA4IE Benchmark.}
\vskip -0.05in
\begin{tabular}{m{1.3cm}m{1.9cm}m{2cm}m{2cm}m{2cm}m{2cm}} \hline
\multicolumn{2}{c}{} & \multicolumn{1}{c}{S} & \multicolumn{1}{c}{M} & \multicolumn{1}{c}{L} & \multicolumn{1}{c}{Total} \\ \hline
\multirow{2}*{SPAN} & \# Docs & \multicolumn{1}{r}{~~~~52898~~~~} & \multicolumn{1}{r}{~~~~29352~~~~} & \multicolumn{1}{r}{~~~~65124~~~~} & \multicolumn{1}{r}{~~~~147374~~~~} \\
~ &\# Triples & \multicolumn{1}{r}{~~~~342361~~~~} & \multicolumn{1}{r}{~~~~195944~~~~} & \multicolumn{1}{r}{~~~~457509~~~~} & \multicolumn{1}{r}{~~~~995814~~~~} \\ \hline
\multirow{4}*{SEQ} & \# Docs & \multicolumn{1}{r}{~~~~52559~~~~} & \multicolumn{1}{r}{~~~~29188~~~~} & \multicolumn{1}{r}{~~~~64385~~~~} & \multicolumn{1}{r}{~~~~146132~~~~} \\
~ &\# Triples & \multicolumn{1}{r}{~~~~341820~~~~} & \multicolumn{1}{r}{~~~~196138~~~~} & \multicolumn{1}{r}{~~~~457033~~~~} & \multicolumn{1}{r}{~~~~994991~~~~} \\
~ &\# Seq-triples & \multicolumn{1}{r}{46521~~~~} & \multicolumn{1}{r}{27176~~~~} & \multicolumn{1}{r}{57507~~~~} & \multicolumn{1}{r}{131204~~~~} \\
~ &\%Seq-triples & \multicolumn{1}{r}{13.61~~~~} & \multicolumn{1}{r}{13.86~~~~} & \multicolumn{1}{r}{12.58~~~~} & \multicolumn{1}{r}{13.19~~~~} \\ \hline
\end{tabular}
\label{table.1}
\vskip -0.1in
\end{table}

\noindent\textbf{Dataset Splitting.}
For comparing the performance on span-triples and seq-triples, we set up two different datasets named QA4IE-SPAN and QA4IE-SEQ. In QA4IE-SPAN, only articles with all span-triples are involved, while in QA4IE-SEQ, articles with seq-triples are also involved.
For studying the influence of the article length as longer articles are normally more difficult to model by LSTMs, we split the articles according to the article length. We name the set of articles with lengths shorter than 400 as S, lengths between 400 and 700 as M, lengths greater than 700 as L. Therefore we obtain 6 different datasets named QA4IE-SPAN-S/M/L and QA4IE-SEQ-S/M/L.
A 5/1/5 splitting of train/dev/test sets is performed.
The detailed statistics of QA4IE benchmark are provided in Table \ref{table.1}.

\begin{table*}[!t]
\centering
\scriptsize
\renewcommand\arraystretch{1.25}
\caption{Comparison between existing IE benchmarks and QA benchmarks. The first two are IE benchmarks and the rest four are QA benchmarks.}
\resizebox{\textwidth}{!}{
\begin{tabular}{|m{2.6cm}<{\centering}|m{3.3cm}<{\centering}|m{1cm}<{\centering}|m{1.8cm}<{\centering}|m{4.4cm}<{\centering}|} \hline
Dataset & Source & \#Docs & \#Triples/Queries & Remarks \\ \hline \hline
\multicolumn{1}{|l|}{\textbf{QA4IE Benchmark}} & \tabincell{c}{Wikipedia / WikiData / DBpedia} & 293K & 2M & automatical generation\\ \hline
\multicolumn{1}{|l|}{Open IE \cite{stanovsky2016creating}} & \tabincell{c}{WSJ / Wikipedia} & 3.2K & 10K & generated from QA-SRL annotations\\ \hline
\multicolumn{1}{|l|}{Zero-Shot Benchmark \cite{levy2017zero}} & \tabincell{c}{Wikipedia / WikiData} & N/A & 30M & sentence level docs, only 120 relation types \\ \hline \hline
\multicolumn{1}{|l|}{WikiReading \cite{hewlett2016wikireading}} & \tabincell{c}{Wikipedia / WikiData} & 4.7M & 18.58M & 11\% errors, 50\% out of document answers \\ \hline
\multicolumn{1}{|l|}{SQuAD \cite{rajpurkar2016squad}} & Wikipedia & 536 & 100K & \tabincell{c}{crowdsourced, span answers only} \\ \hline
\multicolumn{1}{|l|}{CNN/Daily Mail \cite{hermann2015teaching}} & \tabincell{c}{CNN / Daily Mail} & 300K & 1.4M & semi-synthetic cloze-style query \\ \hline
\multicolumn{1}{|l|}{CBT \cite{hill2015goldilocks}} & Children's Book & 688K & 688K & semi-synthetic cloze-style query \\ \hline
\end{tabular}
}
\label{table.2}
\vskip -0.3in
\end{table*}

We further compare our QA4IE benchmark with some existing IE and QA benchmarks in Table \ref{table.2}. One can observe that QA4IE benchmark is much larger than previous IE and QA benchmarks except for WikiReading and Zero-Shot Benchmark. However, as we mentioned at the beginning of Section 2, WikiReading is problematic for IE settings. Besides, Zero-Shot Benchmark is a sentence-level dataset and we have described the disadvantage of ignoring information across sentences at Section 1.1. Thus to our best knowledge, QA4IE benchmark is the largest document level IE benchmark and it can be easily extended if we change our distillation strategy.

\section{Question Answering Model}
\begin{figure}[t]
	\vspace{-10pt}
	\begin{center}
		\centerline{\includegraphics[width=0.8\columnwidth]{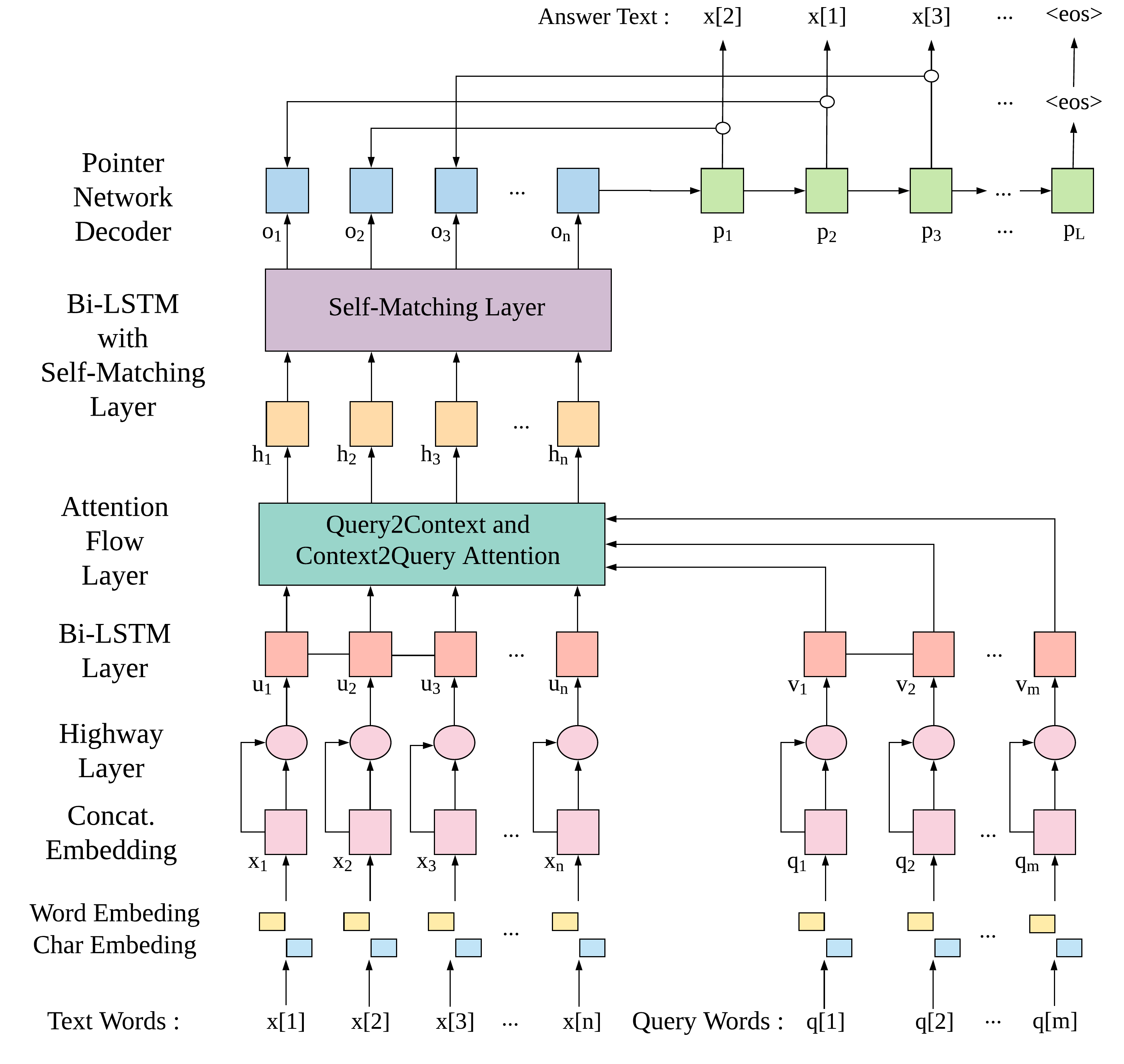}}
		\vskip -0.1in
		\caption{An overview of our QA model.}
		\label{modelfig}
	\end{center}
	\vspace{-35pt}
\end{figure}

In this section, we describe our Question Answering model for IE. The model overview is illustrated in Figure \ref{modelfig}.
The input of our model are the words in the input text $x[1], ... , x[n]$ and query $q[1], ... , q[n]$. We concatenate pre-trained word embeddings from GloVe \cite{pennington2014glove} and character embeddings trained by CharCNN \cite{kim2016character} to represent input words. The $2d$-dimension embedding vectors of input text $x_1, ... , x_n$ and query $q_1, ... , q_n$ are then fed into a Highway Layer \cite{srivastava2015highway} to improve the capability of word embeddings and character embeddings as
\vskip -0.2in
\begin{equation}
\begin{split}
g_t &= {\rm sigmoid}(W_gx_t+b_g) \\
s_t &= {\rm relu  } (W_xx_t+b_x) \\
u_t &= g_t \odot s_t + (1 - g_t) \odot x_t~.
\end{split}
\end{equation}
\vskip -0.08in
Here $W_g, W_x \in \mathbb{R}^{d \times 2d}$ and $b_g, b_x \in \mathbb{R}^d$ are trainable weights, $u_t$ is a $d$-dimension vector.
The function {\rm relu} is the rectified linear units \cite{le2015simple} and $\odot$ is element-wise multiply over two vectors. The same Highway Layer is applied to $q_t$ and produces $v_t$.

Next, $u_t$ and $v_t$ are fed into a Bi-Directional Long Short-Term Memory Network (BiLSTM) \cite{hochreiter1997long} respectively in order to model the temporal interactions between sequence words:
\vskip -0.25in
\begin{equation}
\begin{split}
u_t^{'} &= {\rm BiLSTM}(u^{'}_{t-1},u_t) \\
v_t^{'} &= {\rm BiLSTM}(v^{'}_{t-1},v_t)~.
\end{split}
\end{equation}
\vskip -0.08in

Here we obtain $\mathbf{U} = [u_1^{'}, ... , u_n^{'}] \in \mathbb{R}^{2d \times n}$ and $\mathbf{V} = [v_1^{'}, ... , v_m^{'}] \in \mathbb{R}^{2d \times m}$. Then we feed $\mathbf{U}$ and $\mathbf{V}$ into the attention flow layer \cite{seo2016bidirectional} to model the interactions between the input text and query.
We obtain the $8d$-dimension query-aware context embedding vectors $h_1, ... , h_n$ as the result.

After modeling interactions between the input text and queries, we need to enhance the interactions within the input text words themselves especially for the longer text in IE settings. Therefore, we introduce Self-Matching Layer \cite{wang2017gated} in our model as
\vskip -0.2in
\begin{equation}
\begin{split}
o_t   &= {\rm BiLSTM}(o_{t-1}, [h_t, c_t]) \\
s_j^t &= w^T {\rm tanh}(W_hh_j+\tilde{W_h}h_t)\\
\alpha_i^t &= {\rm exp}(s_i^t)/\Sigma_{j=1}^n{\rm exp}(s_j^t)\\
c_t   &= \Sigma_{i=1}^n\alpha_i^th_i ~.
\end{split}
\end{equation}
\vskip -0.08in
Here $W_h, \tilde{W_h} \in \mathbb{R}^{d \times 8d}$ and $w \in \mathbb{R}^d$ are trainable weights, $[h, c]$ is vector concatenation across row. Besides, $\alpha_i^t$ is the attention weight from the $t^{th}$ word to the $i^{th}$ word and $c_t$ is the enhanced contextual embeddings over the $t^{th}$ word in the input text. We obtain the $2d$-dimension query-aware and self-enhanced embeddings of input text after this step. Finally we feed the embeddings $\mathbf{O} = [o_1, ... , o_n]$ into a Pointer Network \cite{vinyals2015pointer} to decode the answer sequence as
\vskip -0.2in
\begin{equation}
\begin{split}
p_t   &= {\rm LSTM}(p_{t-1}, c_t) \\
s_j^t &= w^T {\rm tanh}(W_oo_j+W_pp_{t-1})\\
\beta_i^t &= {\rm exp}(s_i^t)/\Sigma_{j=1}^n{\rm exp}(s_j^t)\\
c_t   &= \Sigma_{i=1}^n\beta_i^to_i~.
\end{split}
\end{equation}
\vskip -0.08in
The initial state of LSTM $p_0$ is $o_n$. We can then model the probability of the $t^{th}$ token $a^t$ by
\vskip -0.25in
\begin{align}
& {\rm P}(a^t | a^1, ... , a^{t-1}, \mathbf{O}) = (\beta_1^t, \beta_2^t, ... , \beta_n^t, \beta_{n+1}^t) \nonumber \\
& {\rm P}(a^t_i) \triangleq {\rm P}(a^t = i|a^1, ... , a^{t-1}, \mathbf{O}) = \beta_i^t  ~.
\label{eq.5}
\end{align}
\vskip -0.05in
Here $\beta_{n+1}^t$ denotes the probability of generating the ``${\rm eos}$'' symbol since the decoder also needs to determine when to stop.  Therefore, the probability of generating the answer sequence $\textbf{a}$ is as follows
\vskip -0.1in
\begin{equation}
{\rm P}(\textbf{a}|\mathbf{O}) = \prod_t {\rm P}(a^t | a^1, ... , a^{t-1}, \mathbf{O})~.
\end{equation}
\vskip -0.05in
Given the supervision of answer sequence $\mathbf{y} = (y_1, ... , y_L)$, we can write down the loss function of our model as
\vskip -0.15in
\begin{equation}
{\rm L(\theta)} = -\sum_{t=1}^L \log {\rm P} (a^t_{y_t})~.
\label{eq.7}
\end{equation}
\vskip -0.05in
To train our model, we minimize the loss function ${\rm L(\theta)}$ based on training examples.

\section{Experiments}

\subsection{Experimental Setup}
We build our QA4IE benchmark following the steps described in Section 2. In experiments, we train and evaluate our QA models on the corresponding train and test sets while the hyper-parameters are tuned on dev sets. In order to make our experiments more informative, we also evaluate our model on SQuAD dataset \cite{rajpurkar2016squad}.

The preprocessing of our QA4IE benchmark and SQuAD dataset are all performed with the open source code from \cite{seo2016bidirectional}.
We use 100 1D filters with width 5 to construct the CharCNN in our char embedding layer. We set the hidden size $d=100$ for all the hidden states in our model. The optimizer we use is the AdaDelta optimizer \cite{zeiler2012adadelta} with an initial learning rate of 2. A dropout \cite{srivastava2014dropout} rate of 0.2 is applied in all the CNN, LSTM and linear transformation layers in our model during training. For SQuAD dataset and our small sized QA4IE-SPAN/SEQ-S datasets, we set the max length of input texts as 400 and a mini-batch size of 20. For middle sized (and large sized) QA4IE datasets, we set the max length as 700 (800) and batch size as 7 (5). We introduce an early stopping in training process after 10 epochs. Our model is trained on a GTX 1080 Ti GPU and it takes about 14 hours on small sized QA4IE datasets. We implement our model with TensorFlow \cite{abadi2016tensorflow} and optimize the computational expensive LSTM layers with LSTMBlockFusedCell\footnote{https://www.tensorflow.org/api\_docs/python/tf/contrib/rnn/LSTMBlockFusedCell}.

\subsection{Results in QA Settings}
\begin{table}[!t]
\vspace{-5pt}
\centering
\caption{Comparison of QA models on SQuAD datasets. We only include the single model results on the dev set from published papers.}
\vspace{-10pt}
\label{table.3}
\begin{tabular}{m{5cm}c}
 & Dev Set~~ \\ \hline
\emph{Span Model} & EM / F1~~ \\
LR Baseline \cite{rajpurkar2016squad} & 40.0 / 51.0 \\
Match-LSTM \cite{wang2016machine} & 64.1 / 73.9 \\
BiDAF \cite{seo2016bidirectional} & 67.7 / 77.3 \\
R-Net \cite{wang2017gated} & 71.1 / 79.5 \\
MEMEN \cite{pan2017memen} & 71.0 / 80.4 \\
M-Reader+RL \cite{hu2017reinforced} & 72.1 / 81.6 \\
SAN \cite{liu2017stochastic} & \textbf{76.2 / 84.1} \\ \hline
\emph{Sequence Model} &  \\
Match-LSTM (Seq) \cite{wang2016machine}\qquad & 54.4 / 68.2 \\
\textbf{Our Model} & \textbf{61.7 / 72.5} \\ \hline
\end{tabular}
\vspace{-15pt}
\end{table}
We first perform experiments in QA settings to evaluate our QA model on both SQuAD dataset and QA4IE benchmark.
Since our goal is to solve IE, not QA, the motivation of this part of experiments is to evaluate the performance of our model and make a comparison between QA4IE benchmark and existing datasets. Two metrics are introduced in the SQuAD dataset: Exact Match (EM) and F1-score. EM measures the percentage that the model prediction matches one of the ground truth answers exactly while F1-score measures the overlap between the prediction and ground truth answers. Our QA4IE benchmark also adopts these two metrics.

Table \ref{table.3} presents the results of our QA model on SQuAD dataset. Our model outperforms the previous sequence model but is not competitive with span models because it is designed to produce sequence answers in IE settings while baseline span models are designed to produce span answers for SQuAD dataset.

\begin{table}[!t]
\vspace{-5pt}
\scriptsize
\centering
\caption{Comparison of QA models on 6 datasets of our QA4IE benchmark. The BiDAF model cannot work on our SEQ datasets thus the results are N/A.}
\vspace{-5pt}
\begin{tabular}{m{2cm}cccccc}
& SPAN-S~~ & SPAN-M~~ & SPAN-L~~ & SEQ-S~~ & SEQ-M~~ & SEQ-L~~ \\ \hline
Model & EM / F1~~ & EM / F1~~ & EM / F1~~ & EM / F1~~ & EM / F1~~ & EM / F1~~ \\
BiDAF \cite{seo2016bidirectional} & 88.89 / 90.89 & 82.37 / 85.04 & 68.00 / 70.29 &  N/A  &  N/A  &  N/A \\
Match-LSTM \cite{wang2016machine} & 85.88 / 88.21 & 79.19 / 82.05 & 66.87 / 70.44 & 89.60 / 91.95 & 83.57 / 87.40 & 62.64 / 68.98 \\
\textbf{Our Model} & ~~\textbf{91.53 / 93.19}~~ & ~~\textbf{86.04 / 88.65}~~ & ~~\textbf{70.86 / 74.51}~~ & ~~\textbf{91.20 / 93.04}~~ & ~~\textbf{85.52 / 88.43}~~ & ~~\textbf{71.96 / 76.11}~~ \\ \hline
\end{tabular}
\label{table.4}
\end{table}

The comparison between our QA model and two baseline QA models on our QA4IE benchmark is shown in Table \ref{table.4}. For training of both baseline QA models,\footnote{The code of BiDAF is from https://github.com/allenai/bi-att-flow.\\The code of Match-LSTM is from https://github.com/fuhuamosi/MatchLstm.} we use the same configuration of max input length as our model and tune the rest of hyper-parameters on dev sets. Our model outperforms these two baselines on all 6 datasets. The performance is good on S and M datasets but worse for longer documents. As we mentioned in Section 4.1, we set the max input length as 800 and ignore the rest words on L datasets. Actually, there are 11\% of queries with no answers in the first 800 words in our benchmark. Processing longer documents is a tough problem \cite{chen2017reading} and we leave this to our future work.

\begin{table}[!t]
\footnotesize
\centering
\caption{Model ablation on QA4IE-SEQ-S. The first line is our original model and each of the following lines is the original model with a component ablated.}
\vspace{-5pt}
\begin{tabular}{m{4cm}cc} \hline
& EM / F1~~ & ~~~~Training hours~~~~ \\ \hline
\textbf{Our Original Model} & 91.20 / 93.04 & 14 \\
$-$ Char Embedding & 89.78 / 91.76 & 14 \\
$-$ Highway & 90.04 / 91.97 & 14 \\
$-$ Self Matching & 89.55 / 91.60 & 10 \\
$-$ LSTMBlockFusedCell & & 90 \\ \hline
\end{tabular}
\label{table.5}
\vspace{-20pt}
\end{table}

To study the improvement of each component in our model, we present model ablation study results in Table \ref{table.5}. We do not involve Attention Flow Layer and Pointer Network Decoder as they cannot be replaced by other architectures with the model still working.~We can observe that the first three components can effectively improve the performance but Self Matching Layer makes the training more computationally expensive by 40\%. Besides, the LSTMBlockFusedCell works effectively and accelerates the training process by 6 times without influencing the performance.

\subsection{Results in IE Settings}
In this subsection, we put our QA model in the entire pipeline of our QA4IE framework (Figure \ref{framework}) and evaluate the framework in IE settings. Existing IE systems are all free-text based Open IE systems, so we need to manually evaluate the free-text based results in order to compare our model with the baselines. Therefore, we conduct experiments on a small dataset, the dev set of QA4IE-SPAN-S which consists of 4393 documents and 28501 ground truth queries.

Our QA4IE benchmark is based on Wikipedia articles and all the ground truth triples of each article have the same first entity (i.e. the title of the article). Thus, we can directly use the title of the article as the first entity of each triple without performing step 1 (entity recognition) in our framework. Besides, all the ground truth triples in our benchmark are from knowledge base where they are disambiguated and aggregated in the first place, and therefore step 4 (entity linking) is very simple and we do not evaluate it in our experiments.

A major difference between QA settings and IE settings is that in QA settings, each query corresponds to an answer, while in the QA4IE framework, the QA model take a candidate entity-relation (or entity-property) pair as the query and it needs to tell whether an answer to the query can be found in the input text. We can consider the IE settings here as performing step 2 and then step 3 in the QA4IE framework.

In step 2, we need to build a candidate query list for each article in the dataset. Instead of incorporating existing ontology or knowledge base, we use a simple but effective way to build the candidate query list of an article. Since we have a ground truth query list with labeled answers of each article, we can add all the neighboring queries of each ground truth query into the query list. The neighboring queries are defined as two queries that co-occur in the same ground truth query list of any articles in the dataset. We transform the dev set of QA4IE-SPAN-S above by adding neighboring queries into the query list. After this step, the number of queries grows to 426336, and only 28501 of them are ground truth queries labeled with an answer.

\begin{figure}[t]
\vspace{-10pt}
\begin{center}
\centerline{\includegraphics[width=0.7\columnwidth]{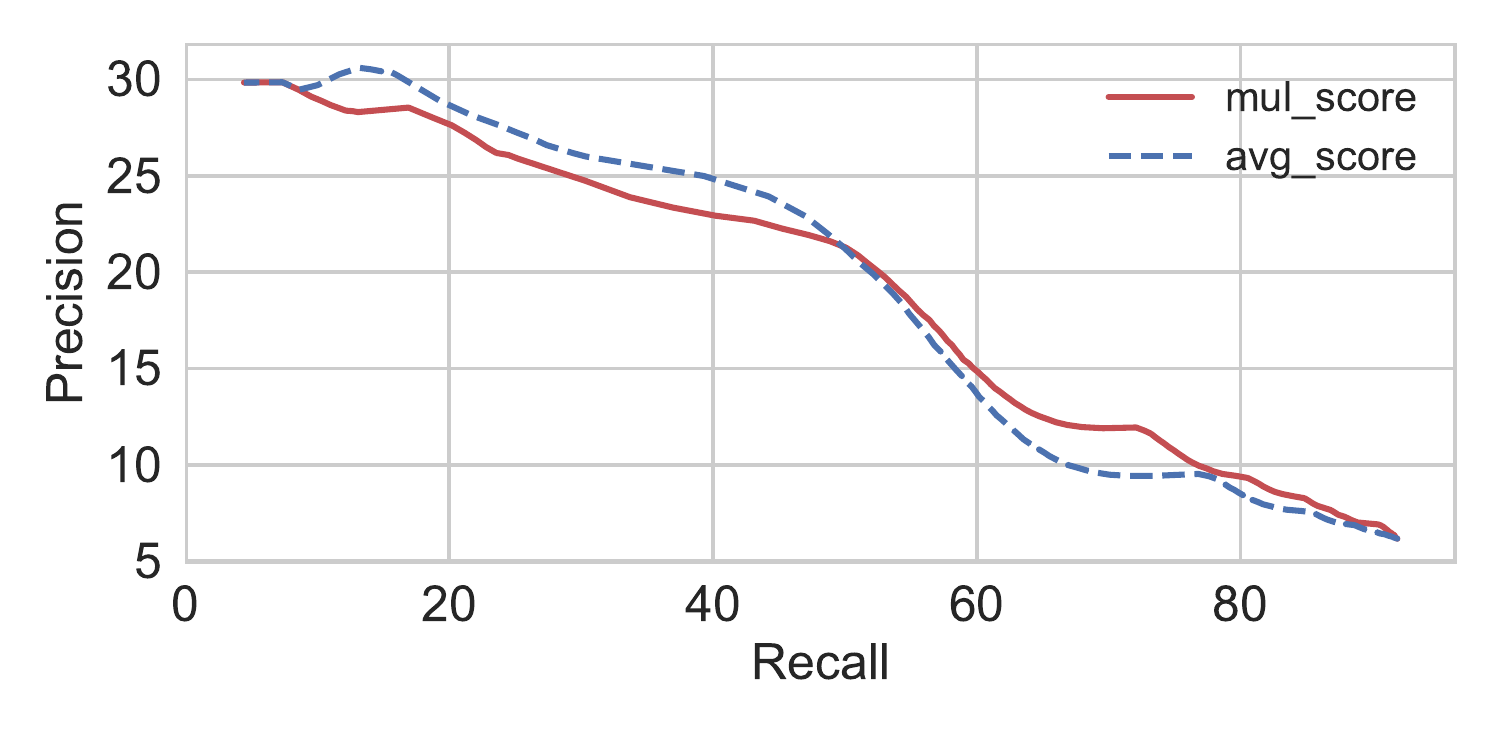}}
\vspace{-15pt}
\caption{Precision-recall curves with two confidence scores on the dev set of QA4IE-SPAN-S.}
\label{pr}
\end{center}
\vspace{-35pt}
\end{figure}

In step 3, we require our QA model to output a confidence score along with the answer to each candidate query. Our QA model produces no answer to a query when the confidence score is less than a threshold $\delta$ or the output is an ``${\rm eos}$'' symbol.
For the answers with a confidence score $\ge \delta$, we evaluate them by the EM measurement with ground truth answers and count the true positive samples in order to calculate the precision and recall under the threshold $\delta$.
Specifically, we try two confidence scores calculated as follows:
\vskip -0.2in
{\small
\begin{equation}
\begin{split}
{\rm Score_{mul}} = \prod_{t=1}^L{\rm P}(a^t_{i_t}),~~~{\rm Score_{avg}} = \sum_{t=1}^L{\rm P}(a^t_{i_t}) / L ~,
\end{split}
\end{equation}}
\vskip -0.15in
\noindent where $(a^1_{i_1}, ... , a^L_{i_L})$ is the answer sequence and ${\rm P}(a^t_i)$ is defined in Eq.~(\ref{eq.5}).${\rm Score_{mul}}$ is equivalent to the training loss in Eq.~(\ref{eq.7}) and ${\rm Score_{avg}}$ takes the answer length into account.

The precision-recall curves of our framework based on the two confidence scores are plotted in Figure \ref{pr}. We can observe that the EM rate we achieve in QA settings is actually the best recall (91.87) in this curve (by setting $\delta = 0$). The best F1-scores of the two curves are 29.97 (precision $= 21.61$, recall $= 48.85$, $\delta = 0.91$) for ${\rm Score_{mul}}$ and 31.05 (precision $= 23.93$, recall $= 44.21$, $\delta = 0.97$) for ${\rm Score_{avg}}$. ${\rm Score_{avg}}$ is better than ${\rm Score_{mul}}$, which suggests that the answer length should be taken into account.

\begin{table}[t]
	\footnotesize
	\centering
    \vspace{-5pt}
	\caption{Results of three Open IE baselines on the dev set of QA4IE-SPAN-S.}
    \vspace{-5pt}
	\begin{tabular}{m{3cm}rrr} \hline
		& ~~~~Open IE 4~~~~ & ~~~~Stanford IE~~~~ & ~~~~ClauseIE~~~~\\ \hline
		\tabincell{l}{\#Extracted Triples} & 32309~~~~ & 120147~~~~ & 75078~~~~\\
		\tabincell{l}{\#After Filtering} & 487~~~~ & 467~~~~ & 554~~~~\\
		\tabincell{l}{\#True Positive} & 403~~~~ & 301~~~~ & 133~~~~\\ \hline
	\end{tabular}
	\label{table.6}
    \vspace{-15pt}
\end{table}

We then evaluate existing IE systems on the dev set of QA4IE-SPAN-S and empirically compare them with our framework. Note that while \cite{levy2017zero} is closely related to our work, we cannot fairly compare our framework with \cite{levy2017zero} because their systems are in the sentence level and require additional negative samples for training. \cite{roth2018neural} is also related to our work, but their dataset and code have not been published yet. Therefore, we choose to evaluate three popular Open IE systems, Open IE 4 \cite{schmitz2012open}, Stanford IE \cite{angeli2015leveraging} and ClauseIE \cite{del2013clausie}.

Since Open IE systems take a single sentence as input and output a set of free-text based triples, we need to find the sentences involving ground truth answers and feed the sentences into the Open IE systems. In the dev set of QA4IE-SPAN-S, there are 28501 queries with 44449 answer locations labeled in the 4393 documents. By feeding the 44449 sentences into the Open IE systems, we obtain a set of extracted triples from each sentence. We calculate the number of true positive samples by first filtering out triples with less than 20\% words overlapping with ground truth answers and then asking two human annotators to verify the remaining triples independently. Since in the experiments, our framework is given the ground-truth first entity of each triple (the title of the corresponding Wikipedia article) while the baseline systems do not have this information, we ask our human annotators to ignore the mistakes on the first entities when evaluating triples produced by the baseline systems to offset this disadvantage. For example, the 3rd case of ClauseIE and the 4th case of Open IE 4 in Table \ref{table.7} are all labeled as correct by our annotators even though the first entities are pronouns.
The two human annotators reached an agreement on 191 out of 195 randomly selected cases.

The evaluation results of the three Open IE baselines are shown in Table \ref{table.6}. We can observe that most of the extracted triples are not related to ground truths and the precision and recall are all very low (around 1\%) although we have already helped the baseline systems locate the sentences containing ground truth answers.

\subsection{Case Study}
\begin{table*}[t]
\scriptsize
\centering
\renewcommand\arraystretch{1.3}
\vskip -0.05in
\caption{Case study of three Open IE baselines and our framework on dev set of QA4IE-SPAN-S, the results of baselines are judged by two human annotators while the results of our framework are measured by Exact Match with ground truth. The triples in red indicate the wrong cases.}
\vspace{-5pt}
\resizebox{\textwidth}{!}{
\begin{tabular}{|m{4.2cm}|m{2.2cm}|m{2.2cm}|m{2.2cm}|m{2.2cm}|m{2.2cm}|} \hline
Input Sentence \centering & Ground Truth Triple \centering & Open IE 4 \centering & Stanford IE \centering
& ClauseIE \centering & \multicolumn{1}{c|}{\textbf{Ours}} \\ \hline
Dieter Kesten was born on 9 June 1914 at Gelsenkirchen. & (Dieter Kesten; date of birth; 9 June 1914) &
(Dieter Kesten; was born; on 9 June 1914 at Gelsenkirchen) & (Dieter Kesten; was born on; 9 June 1914) &
(Dieter Kesten; was born; on 9 June 1914) & (Dieter Kesten; date of birth; 9 June 1914) \\ \hline
Hamilton died on 2 March 1625 at Whitehall, London, from a fever and was buried in the family mausoleum at Hamilton, on 2 September of that year. &
(James Hamilton; date of death; 2 March 1625) &
(Hamilton; died; on 2 March 1625 at Whitehall) &
\makecell[{}{b{2.2cm}}]{{\color{red}{(Hamilton; died on; 2 September)}} \\ (Hamilton; died on; 2 March 1625)} &
{\color{red}{(Hamilton; died on; 2)}}  &
(James Hamilton; date of death; 2 March 1625)\\ \hline
She attended Texas A\&M University, where she swam for the Texas A\&M Aggies swimming and diving team in National Collegiate Athletic Association (NCAA) competition from 2011 to 2014. &
(Breeja Larson; member of sports team; Texas A\&M Aggies) &
{\color{red}{(She; attended; Texas A\&M University)}} &
{\color{red}{(She; attended; M University)}} &
\makecell[{}{b{2.2cm}}]{{\color{red}{(She; attended; Texas A\&M University)}} \\ (she; swam; for the Texas A\&M Aggies swimming and diving team)} &
(Breeja Larson; member of sports team; Texas A\&M Aggies)\\ \hline
His grave and memorial are at Balbeggie Churchyard, St. Martin's, near Perth, Scotland. &
(John Simpson; place of death; St. Martin's) &
(His grave and memorial; are; at Balbeggie Churchyard, St. Martin's, near Perth) &
{\color{red}{(Perth; near Churchyard is; St. Martin's)}} &
{\color{red}{(Balbeggie Churchyard near Perth Scotland; is; St. Martin's)}} &
{\color{red}{(John Simpson; place of death; Balbeggie Churchyard)}} \\ \hline
He served in the British Army and was wounded in World War I. &
(William Dobbie; conflict; World War I) &
{\color{red}{(He; was wounded; in World War I)}} &
{\color{red}{(He; was wounded in; World War I)}} &
{\color{red}{(He; was wounded; in World War I)}} &
(William Dobbie; conflict; World War I) \\ \hline
\end{tabular}
}
\label{table.7}
\vspace{-20pt}
\end{table*}

In this subsection, we perform case studies of IE settings in Table \ref{table.7} to better understand the models and benchmarks. The baseline Open IE systems produce triples by analyzing the subjects, predicates and objects in input sentences, and thus our annotators lower the bar of accepting triples. However, the analysis on semantic roles and parsing trees cannot work very well on complicated input sentences like the 2nd and the 3rd cases. Besides, the baseline systems can hardly solve the last two cases which require inference on input sentences.

Our framework works very well on this dataset with the QA measurements EM $= 91.87$ and F1 $= 93.53$ and the IE measurements can be found in Figure \ref{pr}. Most of the error cases are the fourth case which is acceptable by human annotators. Note that our framework takes the whole document as the input while the baseline systems take the individual sentence as the input, which means the experiment setting is much more difficult for our framework.

\subsection{Human Evaluation on QA4IE Benchmark}
Finally, we perform a human evaluation on our QA4IE benchmark to verify the reliability of former experiments. The evaluation metrics are as follows:

\noindent\textbf{Triple Accuracy} is to check whether each ground truth triple is accurate (one cannot find conflicts between the ground truth triple and the corresponding article) because the ground truth triples from WikiData and DBpedia may be incorrect or incomplete.

\noindent\textbf{Contextual Consistency} is to check whether the context of each answer location is consistent with the corresponding ground truth triple (one can infer from the context to obtain the ground truth triple) because we keep all matched answer locations as ground truths but some of them may be irrelevant with the corresponding triple.

\noindent\textbf{Triple Consistency} is to check whether there is at least one answer location that is contextually consistent for each ground truth triple. It can be calculated by counting the results of Contextual Consistency.

We randomly sample 25 articles respectively from the 6 datasets (in total of 1002 ground truth triples with 2691 labeled answer locations) and let two human annotators label the Triple Accuracy for each ground truth triple and the Contextual Consistency for each answer location. The two human annotators reached an agreement on 131 of 132 randomly selected Triple Accuracy cases and on 229 of 234 randomly selected Contextual Consistency cases. The human evaluation results are shown in Table \ref{table.8}. We can find that the Triple Accuracy and the Triple Consistency is acceptable while the Contextual Consistency still needs to be improved. The Contextual Consistency problem is a weakness of distant supervision, and we leave this to our future work.

\begin{table}[!t]
\vspace{-5pt}
\scriptsize
\centering
\caption{Human evaluation on QA4IE benchmark.}
\vspace{-5pt}
\begin{tabular}{m{2.4cm}ccccccc}
& \multicolumn{1}{c}{SPAN-S} & \multicolumn{1}{c}{SPAN-M} & \multicolumn{1}{c}{SPAN-L} & \multicolumn{1}{c}{SEQ-S} &
\multicolumn{1}{c}{SEQ-M} & \multicolumn{1}{c}{SEQ-L} & \multicolumn{1}{c}{Total} \\ \hline
\multirow{2}*{Triple Accuracy} & 98.8\% & 96.9\% & 98.1\% & 97.1\% & 96.2\% & 97.8\% & 97.5\%\\
~ & ~~161 / 163~~ & ~~154 / 159~~ & ~~159 / 162~~ & ~~170 / 175~~ & ~~152 / 158~~ & ~~181 / 185~~ &~~977 / 1002~\\ \hline
\multirow{2}*{Contextual Consistency} & 78.6\% & 65.1\% & 70.3\% & 75.4\% & 73.9\% & 82.4\% & 74.6\%\\
~ & 195 / 248 & 239 / 367 & 494 / 703 & 230 / 305 & 264 / 357 & 586 / 711 &~2008 / 2691~\\ \hline
\multirow{2}*{Triple Consistency} & 93.3\% & 87.4\% & 91.4\% & 92.0\% & 92.4\% & 92.4\% & 91.5\%\\
~ & 152 / 163 & 139 / 159 & 148 / 162 & 161 / 175 & 146 / 158 & 171 / 185 &~~917 / 1002~\\ \hline
\end{tabular}
\label{table.8}
\vspace{-15pt}
\end{table}

\section{Conclusion}
In this paper, we propose a novel QA based IE framework named QA4IE to address the weaknesses of previous IE solutions. In our framework (Figure \ref{framework}), we divide the complicated IE problem into four steps and show that the step 1, 2 and 4 can be solved well enough by existing work. For the most difficult step 3, we transform it to a QA problem and solve it with our QA model. To train this QA model, we construct a large IE benchmark named QA4IE benchmark that consists of 293K documents and 2 million golden relation triples with 636 different relation types. To our best knowledge, our QA4IE benchmark is the largest document level IE benchmark. We compare our system with existing best IE baseline systems on our QA4IE benchmark and the results show that our system achieves a great improvement over baseline systems.

For the future work, we plan to solve the triples with multiple entities as the second entity, which is excluded from problem scope in this paper. Besides, processing longer documents and improving the quality of our benchmark are all challenging problems as we mentioned previously. We hope this work can provide new thoughts for the area of information extraction.

\section*{Acknowledgements}
W. Zhang is the corresponding author of this paper. The work done by SJTU is sponsored by National Natural Science Foundation of China (61632017, 61702327, 61772333) and Shanghai Sailing Program (17YF1428200).

\bibliography{iswc2018}
\bibliographystyle{splncs04}
\end{document}